\documentclass[11pt,twocolumn]{article}

\usepackage{graphicx}
\usepackage{algorithm,algorithmic}
\usepackage{amsmath}
\usepackage{multirow}
\usepackage{float}
\usepackage{url}
\usepackage[none]{hyphenat}
\usepackage{authblk}

\usepackage{txfonts,balance}

\usepackage[hyphenbreaks]{breakurl}

\usepackage{afterpage}

\begin{document}
\newcounter{cntr1}
\newcounter{cntr2}
\emergencystretch 3em

\title{Circuit simulation using explicit methods: singular matrix issues}

\author{Mahesh~B.~Patil}
\affil{Department of Electrical Engineering, Indian Institute of Technology Bombay}

\maketitle

\begin{abstract}
Some aspects of the ELectrical EXplicit (ELEX) scheme\,\cite{elex1} for
using explicit integration schemes in circuit simulation are discussed.
It is pointed out that the parallel resistor approach, presented earlier
to address singular matrix issues arising in the ELEX scheme, is not
adequately robust for incorporation in a general-purpose simulator
for power electronic circuits. New topology-aware approaches, which
are more robust and efficient compared to the parallel resistor
approach, are presented. Several circuit examples are considered to
illustrate the new approaches.
\end{abstract}

\section{Introduction}
Implicit numerical methods, such as backward Euler
and trapezoidal methods, are commonly used in circuit simulation packages
(e.g., NGSPICE\,\cite{ngspice}, PSIM\,\cite{psim}, PSCAD\,\cite{pscad})
because of their unconditionally stable nature. In the context of power
electronic circuits, explicit methods can also be used for efficient
simulation (e.g., see \cite{plecs}) if the switches are treated as
ideal, i.e., perfect short circuits when closed and perfect open
circuits otherwise. By precomputing
the switching matrices for different switch configurations, a significant
speed-up can be obtained using an explicit scheme, as compared to an
implicit scheme.

As pointed out in \cite{elex1}, a major challenge in implementing an
explicit scheme with ideal switches arises from the singular nature of
the circuit matrix corresponding to specific switch configurations.
One approach to circumvent this problem is to use resistors in parallel
with inductors and switches\,\cite{elex1}. The additional entries that
get created in the circuit matrix because of the parallel resistors
make the matrix non-singular. The parallel resistor approach,
however, is fraught with some difficulties, and it is desirable to
explore alternative approaches.

In this paper, we illustrate the singular matrix issues arising in the
ELEX (ELectrical EXplicit) scheme of \cite{elex1}. We also explain breifly
the parallel resistor approach for addressing these issues and point out
its limitations. We then present alternative ``topology-aware" approaches
which are robust and suitable for implementation in a general circuit
simulation package. In Sec.~\ref{sec_ind}, we consider
a few inductor circuit examples, and in Sec.~\ref{sec_sw}, a few
circuits involving multiple switches.

\section{Inductor circuits}
\label{sec_ind}
\begin{figure}[!ht]
\centering
\scalebox{0.9}{\includegraphics{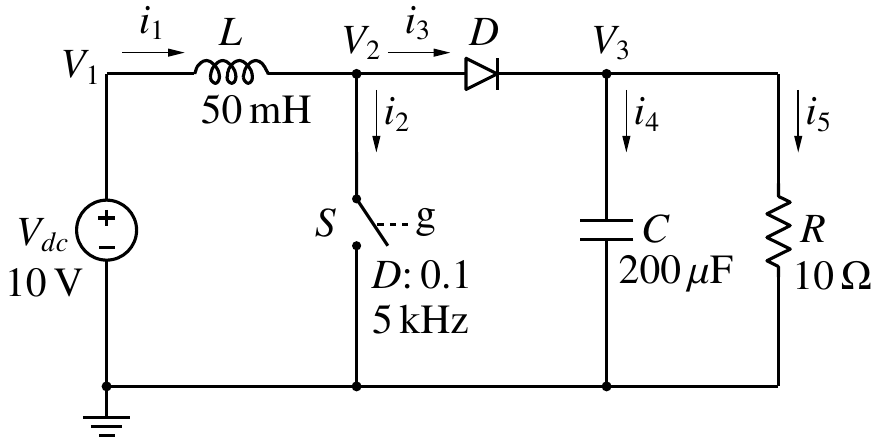}}
\vspace*{-0.2cm}
\caption{Boost converter circuit.}
\label{fig_boost_1}
\end{figure}

\begin{figure}[!ht]
\centering
{\includegraphics[width=0.49\textwidth]{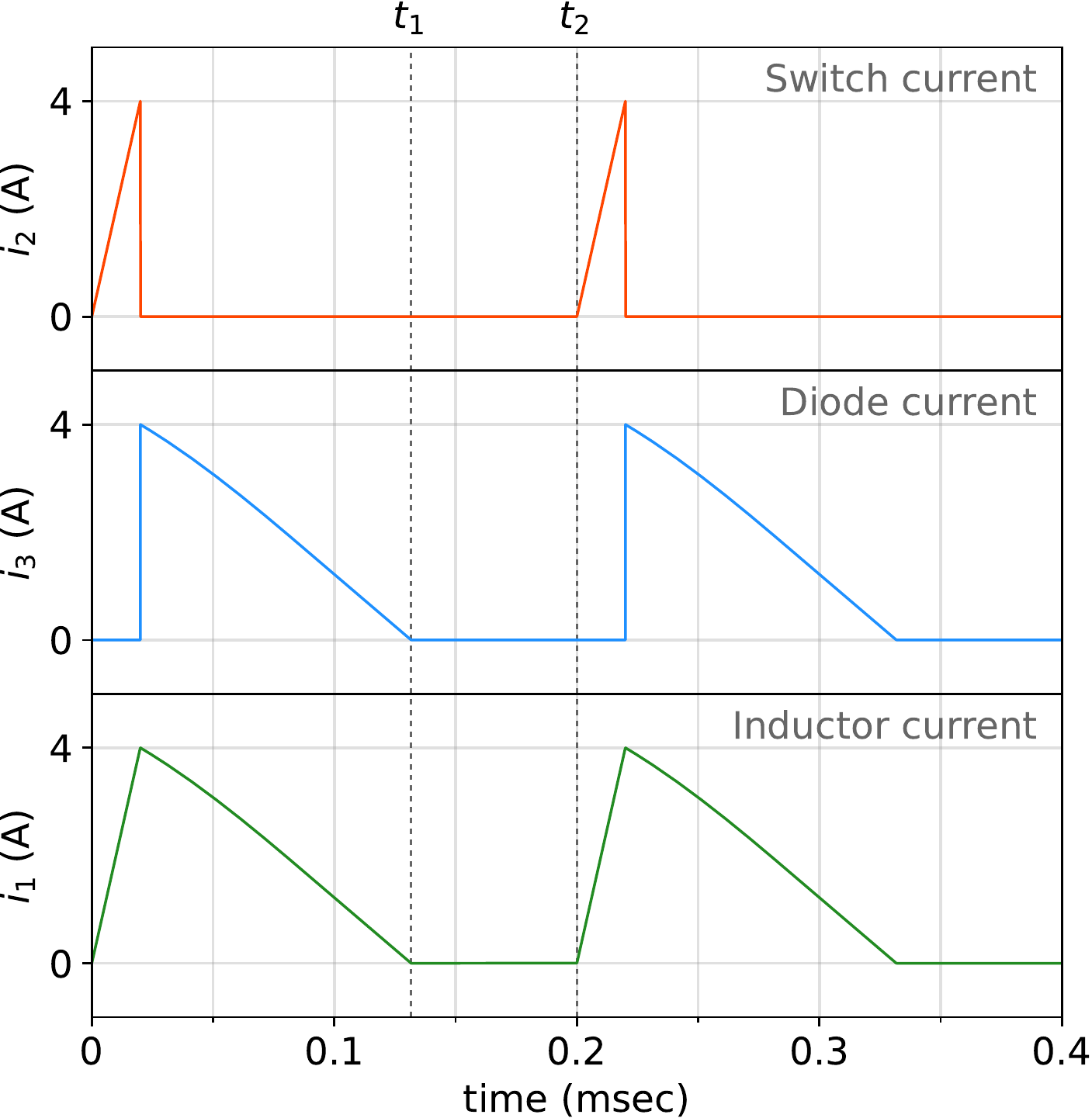}}
\vspace*{-0.7cm}
\caption{Steady-state waveforms for the boost converter circuit
of Fig.~\ref{fig_boost_1}.}
\label{fig_boost_2}
\end{figure}

A boost converter circuit is shown in Fig.~\ref{fig_boost_1} for which
the associated steady-state waveforms are shown in Fig.~\ref{fig_boost_2}.
Let us look at how this circuit will be handled in the ELEX scheme.
For the purpose of illustration, we will consider the forward Euler (FE)
method in the following, although higher-order, variable-step explicit
methods would be preferred in practice\,\cite{elex1},\cite{plecs}.

Suppose we have the solution at time $t_n$ and wish to obtain that at
$t_{n+1}$. In the ELEX scheme, we first update the state variables using
the FE method:
\begin{equation}
V_{C,n+1} = V_{C,n} + \displaystyle\frac{h}{C}\,i_{4,n},
\end{equation}
\begin{equation}
I_{L,n+1} = I_{L,n} + \displaystyle\frac{h}{L}\,\left(V_{1,n}-V_{2,n}\right),
\label{eq_fe_ind}
\end{equation}
where $h \equiv t_{n+1}-t_n$ is the time step. The subscripts $n$ and $n+1$
indicate the values of the concerned variable at $t_n$ and $t_{n+1}$,
respectively. Treating
$V_{C,n+1}$ and
$I_{L,n+1}$
as known quantities, we now write the circuit equations as,
\begin{equation}
V_{1,n+1} = V_{dc},
\label{eq_boost_first}
\end{equation}
\begin{equation}
V_{3,n+1} = V_{C,n+1},
\end{equation}
\begin{equation}
i_{1,n+1} = I_{L,n+1},
\label{eq_boost_ind}
\end{equation}
\begin{equation}
i_{1,n+1} - i_{2,n+1} - i_{3,n+1} = 0,
\label{eq_boost_kcl_1}
\end{equation}
\begin{equation}
i_{3,n+1} - i_{4,n+1} - i_{5,n+1} = 0,
\end{equation}
\begin{equation}
i_{5,n+1} - G\,V_{3,n+1} = 0,
\end{equation}
\begin{equation}
i_{2,n+1} - i_{sw,n+1} = 0,
\end{equation}
\begin{equation}
i_{3,n+1} - i_{d,n+1} = 0,
\end{equation}
where $i_{sw}$ and $i_d$ are the currents through the MOS switch ($S$)
and diode ($D$), respectively, and $G \,$=$\, 1/R$. The equations for
$S$ and $D$ depend on whether they are on or off:
\begin{equation}
\begin{array}{cl}
V_{2,n+1} - V_{3,n+1} = V_{\mathrm{on}} &{\textrm{if}}~D~{\textrm{is on}}, \\
i_{d,n+1} = 0 &{\textrm{if}}~D~{\textrm{is off}},
\end{array}
\end{equation}
\begin{equation}
\begin{array}{cl}
V_{2,n+1} = 0 &{\textrm{if}}~S~{\textrm{is on}}, \\
i_{sw,n+1} = 0 &{\textrm{if}}~S~{\textrm{is off}}.
\end{array}
\label{eq_boost_last}
\end{equation}
Eqs.~\ref{eq_boost_first}-\ref{eq_boost_last} form a linear system of
equations 
(${\bf{A}}\,{\bf{x}} \,$=$\, {\bf{b}}$)
in 10 variables, which can be solved to obtain the solution at $t_{n+1}$.
However, if both $S$ and $D$ are off (see the interval
$t_1 < t < t_2$ in Fig.~\ref{fig_boost_2}), {\bf{A}} becomes singular
because Eq.~\ref{eq_boost_kcl_1} now gives $i_1 \,$=$\, 0$ whereas
Eq.~\ref{eq_boost_ind} is also independently trying to assign a value
to $i_1$ simultaneously.

\begin{figure}[!ht]
\centering
\scalebox{0.9}{\includegraphics{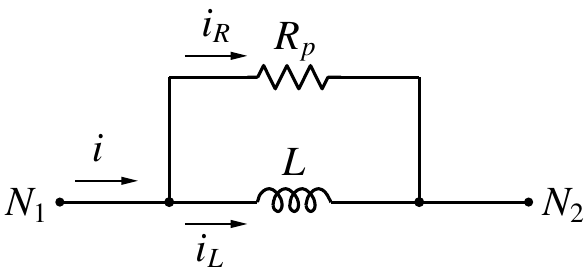}}
\vspace*{-0.2cm}
\caption{Addition of a resistance in parallel with an inductor to address
the singular matrix issue (see text).}
\label{fig_boost_3}
\end{figure}

The above singular matrix situation can be avoided by connecting a
resistance $R_p$ in parallel with the inductor\,\cite{elex1}, as shown
in Fig.~\ref{fig_boost_3}. If the diode turn-off process is handled
carefully, by placing additional time points near the on-to-off transition,
the current through $R_p$ remains small, and the simulation proceeds smoothly.

The $R_p$ approach is attractive from the implementation perspective,
since it is agnostic to the circuit topology. However, we found that it
is not adequately robust in general, because it can introduce small time
constants in the circuit, thus forcing a variable-step simulator to take
extremely small time steps. An alternative way to handle the singular matrix
issue is therefore desirable.

In the following, we propose a ``circuit topology-aware" (CTA) approach
to address singular matrix issues. The CTA approach has two parts:
(a)~element ``stamps"\,\cite{mccalla} which involve equations related
to individual elements, (b)~equations which depend on the circuit topology.
In the proposed scheme, the stamp for the inductor involves two auxiliary
variables, $i_L$ and $i_{Ld}$, corresponding to the inductor current and the derivative
of the inductor current, respectively. It should be noted that $i_L$ merely
{\it{represents}} the inductor current, the actual current being given by
the corresponding branch current. This point will become clear soon.

The element stamp (ES) equations for the inductor are given by,
\begin{equation}
i_L = I_L^{\mathrm{given}},
\end{equation}
\begin{equation}
V_{N1}-V_{N2} - L\,i_{Ld} = 0,
\label{eq_ind_ild}
\end{equation}
where
$I_L^{\mathrm{given}}$
is a known value, computed prior to solving the circuit equations
(see Eq.~\ref{eq_fe_ind}), and $N_1$, $N_2$ are the inductor nodes.

The circuit topology dependent (CTD) equations for the boost converter
of Fig.~\ref{fig_boost_1} can be written using the following observations:
\begin{list}{(\alph{cntr2})}{\usecounter{cntr2}}
 \item
  If there is a conduction path for the inductor, then the inductor
  current is simply equal to the branch current, which satisfies KCL.
 \item
  If there is no conduction path for the inductor, the inductor current
  is zero, a constant, and therefore we expect the inductor voltage
  to become zero. This condition can be forced by making $i_{Ld}$
  equal to zero (see Eq.~\ref{eq_ind_ild}).
\end{list}
With these considerations, we can now write the CTD equations:
\begin{equation}
\begin{array}{cl}
i_{Ld} = 0 &{\textrm{if}}~S~{\textrm{and}}~D~{\textrm{are both off}}, \\
i_1 - i_L = 0 &{\textrm{otherwise}}.
\end{array}
\end{equation}
Combining the ES and CTD equations, we now have the following set of equations
for the boost converter.
(For convenience, we have dropped the subscript $n+1$ from the variables.
For example, $V_1$ in the following should be taken to mean $V_{1,n+1}$, and so on.)
\begin{equation}
V_1 = V_{dc},
\end{equation}
\begin{equation}
V_3 = V_{C,n+1},
\end{equation}
\begin{equation}
i_L = I_{L,n+1},
\end{equation}
\begin{equation}
V_1 - V_2 - L\,i_{Ld} = 0,
\end{equation}
\begin{equation}
i_1 - i_2 - i_3 = 0,
\end{equation}
\begin{equation}
i_3 - i_4 - i_5 = 0,
\end{equation}
\begin{equation}
i_5 - G\,V_3 = 0,
\end{equation}
\begin{equation}
i_2 - i_{sw} = 0,
\end{equation}
\begin{equation}
i_3 - i_d = 0,
\end{equation}
\begin{equation}
\begin{array}{cl}
V_2 - V_3 = V_{\mathrm{on}} &{\textrm{if}}~D~{\textrm{is on}}, \\
i_d = 0 &{\textrm{if}}~D~{\textrm{is off}},
\end{array}
\end{equation}
\begin{equation}
\begin{array}{cl}
V_2 = 0 &{\textrm{if}}~S~{\textrm{is on}}, \\
i_{sw} = 0 &{\textrm{if}}~S~{\textrm{is off}}.
\end{array}
\end{equation}
\begin{equation}
\begin{array}{cl}
i_{Ld} = 0 &{\textrm{if}}~S~{\textrm{and}}~D~{\textrm{are both off}}, \\
i_1 - i_L = 0 &{\textrm{otherwise}}.
\end{array}
\end{equation}
We now have a system of equations with 12 variables
($V_1$, $V_2$, $V_3$, $i_1$, $i_2$, $i_3$, $i_4$, $i_5$,
$i_d$, $i_{sw}$, $i_L$, $i_{Ld}$).
When both $S$ and $D$ are off,
$i_1 \,$=$\, i_L$
is not used; it gets replaced with
$V_1 - V_2 \,$=$\, L\,i_{Ld} = 0$, and the
matrix is not singular any more.

The other aspects of the ELEX scheme, such as consistency check
for the switches and time step control, remain the same as before\,\cite{elex1}.
We have verified that the proposed CTA equations, when implemented in the
ELEX-RKF scheme of \cite{elex1}, gives the results expected for the boost
converter.

We now look at a few additional examples and illustrate the construction of
the CTD equations in each case.
\subsection{Inductor circuits: example 1}
\label{sec_ind_ex_1}
\begin{figure}[!ht]
\centering
\scalebox{0.9}{\includegraphics{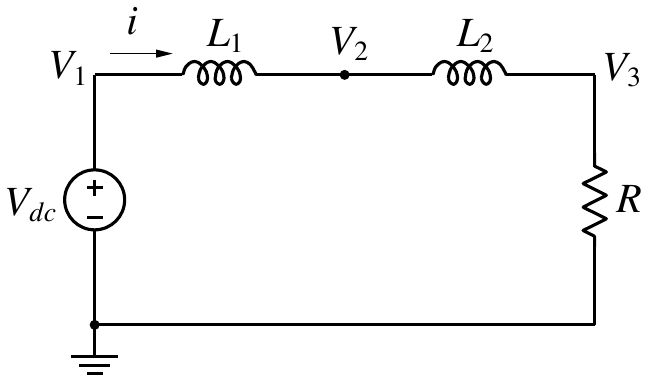}}
\vspace*{-0.2cm}
\caption{Inductor circuit example.}
\label{fig_ind_ex_1}
\end{figure}

Consider the circuit shown in Fig.~\ref{fig_ind_ex_1}. The ES equations
in this case are given by,
\begin{equation}
V_1 = V_{dc},
\end{equation}
\begin{equation}
i - G\,V_3 = 0,
\end{equation}
\begin{equation}
i_{L1} = I_{L1,n+1},
\end{equation}
\begin{equation}
V_1 - V_2 -L_1\,i_{L1d} = 0,
\end{equation}
\begin{equation}
i_{L2} = I_{L2,n+1},
\end{equation}
\begin{equation}
V_2 - V_3 -L_2\,i_{L2d} = 0,
\end{equation}
where
$I_{L1,n+1}$,
$I_{L2,n+1}$
are constants. If the FE method is used, they are given by,
\begin{equation}
I_{L1,n+1} = I_{L1,n} + \displaystyle\frac{h}{L_1}\,\left(V_{1,n}-V_{2,n}\right),
\nonumber
\end{equation}
\begin{equation}
I_{L2,n+1} = I_{L2,n} + \displaystyle\frac{h}{L_2}\,\left(V_{2,n}-V_{3,n}\right).
\nonumber
\end{equation}

To construct the CTD equations for this circuit, we observe the following:
\begin{list}{(\alph{cntr2})}{\usecounter{cntr2}}
 \item
  Since the two inductor currents are identical, we can equate one of the two
  auxiliary variables (i.e., $i_{L1}$ or $i_{L2}$) to
  the branch current $i$. Equating each of them to $i$ would result in a
  singular matrix, which we want to avoid.
 \item
  We expect the two inductor voltages $V_{L1}$ and $V_{L2}$
  to be related by
  $\displaystyle\frac{V_{L1}}{V_{L2}} \,$=$\, \displaystyle\frac{L_1}{L_2}$
  since the derivatives
  $\displaystyle\frac{di_{L1}}{dt}$ and
  $\displaystyle\frac{di_{L2}}{dt}$
  are identical. This condition can be forced by simply equating
  $i_{L1d}$ and $i_{L2d}$.
\end{list}
The CTD equations can now be written as
\begin{equation}
i - i_{L1} = 0,
\end{equation}
\begin{equation}
i_{L1d} - i_{L2d} = 0,
\end{equation}
thus completing the system of equations with 8 variables ($V_1$,
$V_2$, $V_3$, $i$, $i_{L1}$, $i_{L1d}$, $i_{L2}$, $i_{L2d}$).

\subsection{Inductor circuits: example 2}
\label{sec_ind_ex_2}
\begin{figure}[!ht]
\centering
\scalebox{0.9}{\includegraphics{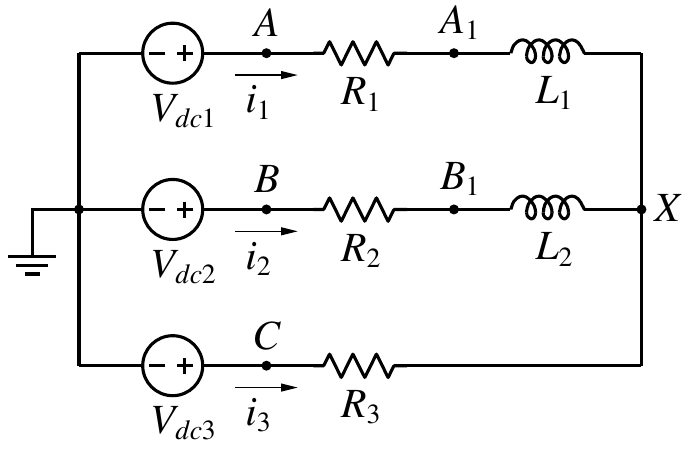}}
\vspace*{-0.2cm}
\caption{Inductor circuit example.}
\label{fig_ind_ex_2}
\end{figure}

We now consider a circuit in which two branches with one inductor each
and one branch without any inductor
share a common node (see Fig.~\ref{fig_ind_ex_2}). The ES equations
are given by,
\begin{equation}
V_A = V_{dc1},
\end{equation}
\begin{equation}
V_B = V_{dc2},
\end{equation}
\begin{equation}
V_C = V_{dc3},
\end{equation}
\begin{equation}
i_1 - G_1\,\left(V_A-V_{A1}\right) = 0,
\end{equation}
\begin{equation}
i_2 - G_2\,\left(V_B-V_{B1}\right) = 0,
\end{equation}
\begin{equation}
i_3 - G_3\,\left(V_C-V_X\right) = 0,
\end{equation}
\begin{equation}
i_{L1} = I_{L1,n+1},
\end{equation}
\begin{equation}
V_{A1} - V_X -L_1\,i_{L1d} = 0,
\end{equation}
\begin{equation}
i_{L2} = I_{L2,n+1},
\end{equation}
\begin{equation}
V_{B1} - V_X -L_2\,i_{L2d} = 0,
\end{equation}
where
$I_{L1,n+1}$,
$I_{L2,n+1}$
are constants. If the FE method is used, they are given by,
\begin{equation}
I_{L1,n+1} = I_{L1,n} + \displaystyle\frac{h}{L_1}\,\left(V_{A1,n}-V_{X,n}\right),
\nonumber
\end{equation}
\begin{equation}
I_{L2,n+1} = I_{L2,n} + \displaystyle\frac{h}{L_2}\,\left(V_{B1,n}-V_{X,n}\right).
\nonumber
\end{equation}
We now come to the CTD equations.
Since branches 1 and 2 have only one inductor each, we can equate the
branch current to the corresponding $i_L$ and get
\begin{equation}
i_1 - i_{L1} = 0,
\end{equation}
\begin{equation}
i_2 - i_{L2} = 0,
\end{equation}
and finally, KCL at node $X$ gives
\begin{equation}
i_1 + i_2 + i_3 = 0,
\end{equation}
giving 13 equations in 13 variables ($V_A$, $V_B$, $V_C$, $V_{A1}$, $V_{B1}$, $V_X$,
$i_1$, $i_2$, $i_3$, $i_{L1}$, $i_{L1d}$, $i_{L2}$, $i_{L2d}$).

\subsection{Inductor circuits: example 3}
\label{sec_ind_ex_3}
\begin{figure}[!ht]
\centering
\scalebox{0.9}{\includegraphics{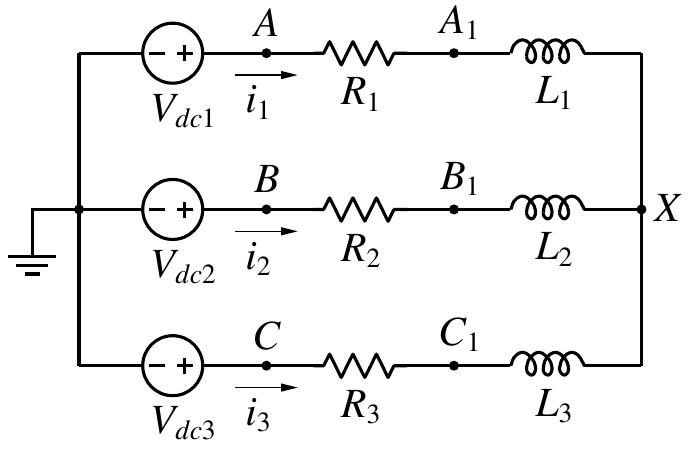}}
\vspace*{-0.2cm}
\caption{Inductor circuit example.}
\label{fig_ind_ex_3}
\end{figure}

Consider the circuit shown in Fig.~\ref{fig_ind_ex_3}, which is
similar to the previous circuit (Fig.~\ref{fig_ind_ex_2}) except
that branch 3 also has an inductor. The ES equations are given by,
\begin{equation}
V_A = V_{dc1},
\end{equation}
\begin{equation}
V_B = V_{dc2},
\end{equation}
\begin{equation}
V_C = V_{dc3},
\end{equation}
\begin{equation}
i_1 - G_1\,\left(V_A-V_{A1}\right) = 0,
\end{equation}
\begin{equation}
i_2 - G_2\,\left(V_B-V_{B1}\right) = 0,
\end{equation}
\begin{equation}
i_3 - G_3\,\left(V_C-V_{C1}\right) = 0,
\end{equation}
\begin{equation}
i_{L1} = I_{L1,n+1},
\end{equation}
\begin{equation}
V_{A1} - V_X -L_1\,i_{L1d} = 0,
\end{equation}
\begin{equation}
i_{L2} = I_{L2,n+1},
\end{equation}
\begin{equation}
V_{B1} - V_X -L_2\,i_{L2d} = 0,
\end{equation}
\begin{equation}
i_{L3} = I_{L3,n+1},
\end{equation}
\begin{equation}
V_{C1} - V_X -L_3\,i_{L3d} = 0,
\end{equation}
where
$I_{L1,n+1}$,
$I_{L2,n+1}$,
$I_{L3,n+1}$
are constants which, with the FE method, are given by,
\begin{equation}
I_{L1,n+1} = I_{L1,n} + \displaystyle\frac{h}{L_1}\,\left(V_{A1,n}-V_{X,n}\right),
\nonumber
\end{equation}
\begin{equation}
I_{L2,n+1} = I_{L2,n} + \displaystyle\frac{h}{L_2}\,\left(V_{B1,n}-V_{X,n}\right),
\nonumber
\end{equation}
\begin{equation}
I_{L3,n+1} = I_{L3,n} + \displaystyle\frac{h}{L_3}\,\left(V_{C1,n}-V_{X,n}\right).
\nonumber
\end{equation}
To write the CTD equations, we first equate each branch current to the $i_L$ variable
in that branch:
\begin{equation}
i_1 - i_{L1} = 0,
\label{eq_ind_3_kcl_1}
\end{equation}
\begin{equation}
i_2 - i_{L2} = 0,
\label{eq_ind_3_kcl_2}
\end{equation}
\begin{equation}
i_3 - i_{L3} = 0.
\label{eq_ind_3_kcl_3}
\end{equation}
At this stage, we have 15 equations in 16 variables ($V_A$, $V_B$, $V_C$,
$V_{A1}$, $V_{B1}$, $V_{C1}$, $V_X$,
$i_1$, $i_2$, $i_3$,
$i_{L1}$, $i_{L1d}$, $i_{L2}$, $i_{L2d}$, $i_{L3}$, $i_{L3d}$),
and one more equation is required.
Although the topology for this circuit is similar to the previous circuit, there is a
significant difference: in this case, all three branches connected to the common node
$X$ have inductors. The equation we are looking for cannot be KCL at node $X$~-- that
would cause a singular matrix error because
Eqs.~\ref{eq_ind_3_kcl_1}-\ref{eq_ind_3_kcl_3} are already specifying
$i_1$,
$i_2$,
$i_3$.
Instead, we can use the fact that the derivatives of the three inductor currents must
also add to zero, i.e.,
\begin{equation}
i_{L1d} + i_{L2d} + i_{L3d} = 0.
\end{equation}

From the above examples, it is clear that implementation of the CTA approach in
a circuit simulator is somewhat tricky since it requires several special cases to
be considered. However, if suitable rules are formulated for constructing the CTD 
equations, this task can be simplified. Based on the above discussion, we can
make up the following rules.
\begin{list}{\arabic{cntr1}.}{\usecounter{cntr1}}
 \item
  If a branch has multiple inductors (i.e., inductors connected in series),
  equate the branch current to only one of the $i_L$ variables in that branch.
  Make the $i_{Ld}$ variables equal for all inductors in that branch. For example,
  if the branch has three inductors, we write
  $i_{L2d} \,$=$\, i_{L1d}$,
  $i_{L3d} \,$=$\, i_{L1d}$.
 \item
  If two or more branches share a common node, use the following rules.
  \begin{list}{(\alph{cntr2})}{\usecounter{cntr2}}
   \item
    If one of the branches does not have an inductor, write KCL at the common
    node.
   \item
    If each branch has one or more inductors, take one $i_{Ld}$ variable from each
    branch and write
    $\displaystyle\sum_{k} i_{Ld}^{(k)} \,$=$\, 0$
    where $k$ is the branch index.
  \end{list}
 \item
  Consider a node at which an inductor branch is incident. If all other branches
  incident on this node are open (because they have open switches), then make
  $i_{Ld} \,$=$\, 0$; else, write KCL at that node.
\end{list}
The list of rules presented above is not exhaustive, but it does give an idea
of the considerations that must go into devising a general set of rules for the CTD
equations.

\section{Switch circuits}
\label{sec_sw}
\begin{figure}[!ht]
\centering
\scalebox{0.9}{\includegraphics{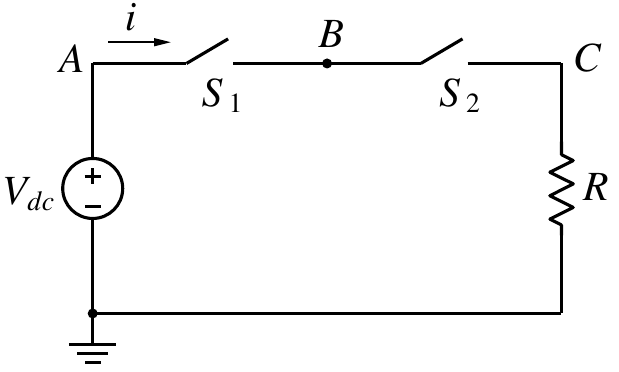}}
\vspace*{-0.2cm}
\caption{Circuit with switches in series.}
\label{fig_sw_series}
\end{figure}

In the ELEX scheme, an on switch is represented by a short circuit (or a constant
voltage drop $V_{\mathrm{on}}$), and an off switch by an open circuit.
Let us consider a simple example (see Fig.~\ref{fig_sw_series})
in which two switches are connected in series. Note that, for
simplicity, we have not considered any capacitors or inductors here,
which means that the solution would be obtained by solving the circuit
equations only once.

If both $S_1$ and $S_2$ are off, we have
$i_{sw}^{(1)} \,$=$\, 0$,
$i_{sw}^{(2)} \,$=$\, 0$,
$i_{sw}^{(1)}$, $i_{sw}^{(2)}$ being the switch currents.
In this case, node $B$ gets isolated, and there is no way to determine $V_B$.
In other words, the circuit matrix would turn out to be singular. In practice,
we would expect $V_{dc}$ to get divided equally between
$S_1$ and $S_2$ (if the two switches are identical), thus giving
$V_B \,$=$\, V_{dc}/2$.

\begin{figure}[!ht]
\centering
\scalebox{0.9}{\includegraphics{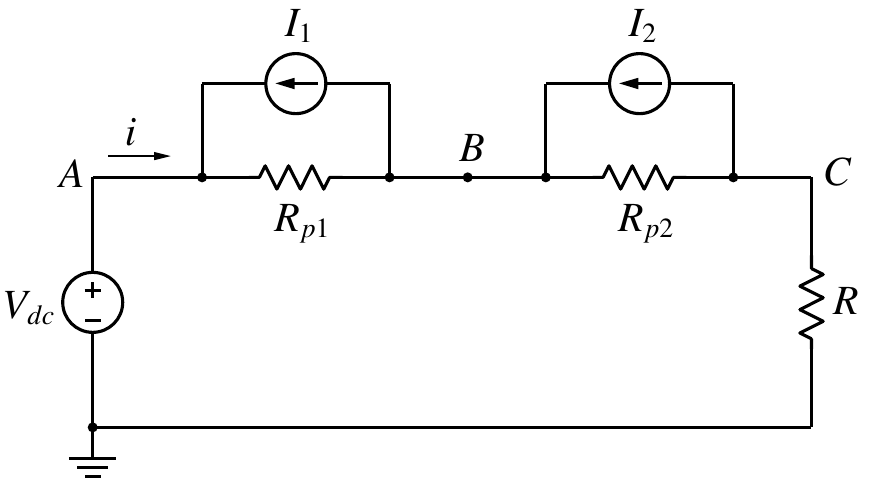}}
\vspace*{-0.2cm}
\caption{Circuit of Fig.~\ref{fig_sw_series} with each switch replaced by
a resistor-current source combination.}
\label{fig_sw_series_rp}
\end{figure}

One approach to obtain the desired solution\,\cite{elex1} is to replace each
off switch with a resistor-current source combination, as shown in
Fig.~\ref{fig_sw_series_rp}. The resistances $R_{p1}$ and $R_{p2}$ can be
chosen to be large.
The circuit equations can now be written as,
\begin{equation}
V_A = V_{dc},
\end{equation}
\begin{equation}
i - G_{p1}\,\left(V_A - V_B\right) = -I_1,
\label{eq_sw_rp_1}
\end{equation}
\begin{equation}
i - G_{p2}\,\left(V_B - V_C\right) = -I_2,
\label{eq_sw_rp_2}
\end{equation}
\begin{equation}
i - G\,V_c = 0,
\end{equation}
where
$G \,$=$\, 1/R$,
$G_{p1} \,$=$\, 1/R_{p1}$,
$G_{p2} \,$=$\, 1/R_{p2}$.
The idea is to change
$I_1$ from $0$ to $G_{p1}\,\left(V_A - V_B\right)$
(and similarly, $I_2$ from $0$ to $G_{p2}\,\left(V_B - V_C\right)$)
in a few iterations. In other words, we start with
$I_1 \,$=$\, 0$,
$I_2 \,$=$\, 0$,
compute the solution, use the values of $V_A$, $V_B$, $V_C$ to update
$I_1$ as
$\displaystyle\frac{k-1}{k_{\mathrm{max}}-1}\,G_{p1}\,(V_A-V_B)$
($k$ being the iteration number), and so on up to $k \,$=$\, k_{\mathrm{max}}$.
$I_2$ is also similarly updated in each iteration. If this process converges,
$G_{p1}\,\left(V_A - V_B\right)$ cancels with $I_1$ in Eq.~\ref{eq_sw_rp_1},
and the equation reduces to $i \,$=$\, 0$. This amounts to making
$R_{p1}$ (and similarly, $R_{p2}$) infinite. Note that the circuit matrix
remains constant and non-singular throughout.

The above approach, though acceptable for some circuits, is not suitable in
general for the following reasons.
\begin{list}{(\alph{cntr2})}{\usecounter{cntr2}}
\item
 It increases the computation time.
\item
 Convergence rate depends on circuit parameters, which has a negative impact
 on robustness of the procedure.
\item
 If there are several switches in the circuit, convergence could be difficult
 (or even impossible) to achieve.
\end{list}

We now present a CTA approach to address the singular matrix issue arising
for the circuit of Fig.~\ref{fig_sw_series} when both $S_1$ and $S_2$ are off.
As in Sec.~\ref{sec_ind}, we divide the equations into two categories: ES and
CTD. A switch has the following stamp:
\begin{equation}
\begin{array}{cl}
i_{sw} = 0 &{\textrm{if}}~S~{\textrm{is off}}, \\
V_{sw} = V_{\mathrm{on}} &{\textrm{if}}~S~{\textrm{is on}},
\label{eq_sw_1}
\end{array}
\end{equation}
where $V_{\mathrm{on}}$ is the voltage drop across the switch when it is
conducting. For the switches in Fig.~\ref{fig_sw_series}, we will consider
$V_{\mathrm{on}}$  to be zero. Note that $V_{sw}$ and $i_{sw}$ in Eq.~\ref{eq_sw_1}
represent the switch voltage and switch current, respectively. They need to be
related to the circuit currents and voltages using the CTD equations, as we will
see in the following.

The ES equations for the circuit of Fig.~\ref{fig_sw_series}, with both $S_1$
and $S_2$ off, are given by,
\begin{equation}
V_A = V_{dc},
\label{eq_sw_series_first}
\end{equation}
\begin{equation}
i_{sw}^{(1)} = 0,
\end{equation}
\begin{equation}
i_{sw}^{(2)} = 0,
\end{equation}
\begin{equation}
i - G\,V_C = 0.
\end{equation}

In order to write the CTD equations, we observe the following.
\begin{list}{(\alph{cntr2})}{\usecounter{cntr2}}
 \item
  The branch current must be zero since the branch has one or more switches which
  are not conducting. Equivalently, we could equate the branch current to
  $i_{sw}^{(1)}$ or $i_{sw}^{(2)}$.
 \item
  We expect $S_1$ and $S_2$ to share the voltage drop equally, assuming that they
  are identical switches.
\end{list}
We can now write the CTD equations as
\begin{equation}
V_A - V_B - V_{sw}^{(1)} = 0,
\end{equation}
\begin{equation}
V_b - V_C - V_{sw}^{(2)} = 0,
\end{equation}
\begin{equation}
V_{sw}^{(1)} - V_{sw}^{(2)} = 0,
\end{equation}
\begin{equation}
i = 0.
\label{eq_sw_series_last}
\end{equation}
Eqs.~\ref{eq_sw_series_first}-\ref{eq_sw_series_last} form a linear system in 8
variables ($V_A$, $V_B$, $V_C$, $i$, $i_{sw}^{(1)}$, $i_{sw}^{(2)}$, $V_{sw}^{(1)}$,
$V_{sw}^{(2)}$), which can be solved to yield the desired solution, viz.,
$V_A \,$=$\, V_{dc}$,
$V_B \,$=$\, V_{dc}/2$,
$V_C \,$=$\, 0$,
$i \,$=$\, 0$,
$i_{sw}^{(1)} \,$=$\, i_{sw}^{(2)} \,$=$\, 0$,
$V_{sw}^{(1)} \,$=$\, V_{sw}^{(2)} \,$=$\, V_{dc}/2$.

We will now look at two test cases which will help in formulating some
general rules for CTD equations for circuit with switches.

\subsection{Switch circuits: example 1}
\label{sec_sw_ex_1}
\begin{figure}[!ht]
\centering
\scalebox{0.9}{\includegraphics{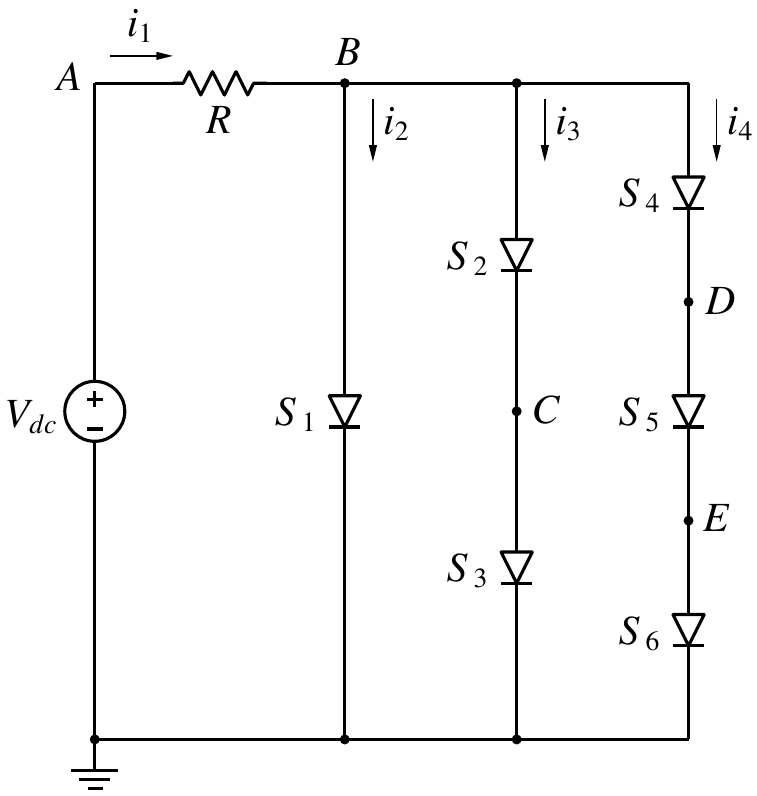}}
\vspace*{-0.2cm}
\caption{Switch circuit example.}
\label{fig_sw_ex_1}
\end{figure}

Consider the circuit shown in Fig.~\ref{fig_sw_ex_1} in which the diodes have on-state
voltage drops
$V_{\mathrm{on}}^{(1)}$,
$V_{\mathrm{on}}^{(2)}$, etc.
The ES equations are given by,
\begin{equation}
V_A = V_{dc},
\end{equation}
\begin{equation}
i - G\,\left(V_A - V_B\right) = 0,
\end{equation}
\begin{equation}
\begin{array}{cl}
i_{sw}^{(k)} = 0 &{\textrm{if}}~S_k~{\textrm{is off}}, \\
V_{sw}^{(k)} = V_{\mathrm{on}}^{(k)} &{\textrm{if}}~S_k~{\textrm{is on}}.
\end{array}
\label{eq_sw_ex_1_stamp}
\end{equation}
Note that Eq.~\ref{eq_sw_ex_1_stamp} represents six equations, one for each switch.
The number of ES equations is therefore eight. There are 21 variables, viz.,
$V_A$ to $V_E$, $i_1$ to $i_4$,
$i_{sw}^{(1)}$ to $i_{sw}^{(6)}$, and
$V_{sw}^{(1)}$ to $V_{sw}^{(6)}$. We would therefore expect 13 equations to come from
the CTD set. Some of the CTD equations are valid irrespective of the on/off status of the
switches. We will refer to these equations as CTD-Constant or CTDC equations. The
remaining CTD equations would vary, depending on which switches are on, and we will
refer to them as CTD-Variable or CTDV equations.

The CTDC equations are given by,
\begin{equation}
i_1 - i_2 - i_3 - i_4 = 0,
\label{eq_sw_ex_1_1}
\end{equation}
\begin{equation}
i_1 - i_{sw}^{(1)} = 0,
\label{eq_sw_ex_1_2}
\end{equation}
\begin{equation}
i_2 - i_{sw}^{(2)} = 0,
\label{eq_sw_ex_1_3}
\end{equation}
\begin{equation}
i_3 - i_{sw}^{(4)} = 0,
\label{eq_sw_ex_1_4}
\end{equation}
where Eq.~\ref{eq_sw_ex_1_1} comes from KCL at node $B$.
Eqs.~\ref{eq_sw_ex_1_2}-\ref{eq_sw_ex_1_4} are obtained by equating each branch
current to one of the $i_{sw}$ variables in that branch.

To formulate the CTDV equations, we categorise each switch branch (i.e., a branch
containing one or more switches) as an ON branch if all switches in that branch are
on and as an OFF branch otherwise. To be specific, let us take
$S_2$, $S_3$, $S_5$ to be on and the others off. In this case, branch 3 (carrying
current $i_3$) is ON while branches 2 and 4 are OFF.

For each OFF branch, we relate
$V_{sw}^{(k)}$
to the terminal voltages for each switch in that branch, and obtain
\begin{equation}
V_B - 0 - V_{sw}^{(1)} = 0,
\end{equation}
\begin{equation}
V_B - V_D - V_{sw}^{(4)} = 0,
\end{equation}
\begin{equation}
V_D - V_E - V_{sw}^{(5)} = 0,
\end{equation}
\begin{equation}
V_E - 0 - V_{sw}^{(6)} = 0.
\end{equation}

Next, for each OFF branch, we distribute the ``reverse bias" equally between
the off switches, which gives
\begin{equation}
(V_{sw}^{(4)} - V_{\mathrm{on}}^{(4)}) -
(V_{sw}^{(6)} - V_{\mathrm{on}}^{(6)}) = 0.
\end{equation}
We should note that, if $D_4$ and $D_6$ were to be pointing in opposite directions,
the equation would be
\begin{equation}
(V_{sw}^{(4)} - V_{\mathrm{on}}^{(4)}) +
(V_{sw}^{(6)} - V_{\mathrm{on}}^{(6)}) = 0.
\nonumber
\end{equation}

If there are any on switches in an OFF branch, we make the corresponding
$i_{sw}^{(k)}$ equal to zero. For branch 4, therefore, we have
\begin{equation}
i_{sw}^{(5)} = 0.
\end{equation}

For each ON branch, if there are multiple switches, we equate their $i_{sw}$
variables. In our example, only branch 2 is ON, and it contains $S_2$ and $S_3$,
leading to
\begin{equation}
i_{sw}^{(3)} - i_{sw}^{(2)} = 0.
\end{equation}

Finally, we relate
$V_{sw}^{(k)}$ to the terminal voltages for each switch in the ON branch to get
\begin{equation}
V_B - V_C - V_{sw}^{(2)} = 0,
\end{equation}
\begin{equation}
V_C - 0 - V_{sw}^{(3)} = 0.
\end{equation}

Solving the above sets of equations (i.e., the ES equations and the CTD equations
together) gives us the solution for the condition we have assumed, viz.,
$S_2$, $S_3$, $S_5$ on and the other switches off. The solution so obtained, however,
may not be the solution we are interested in since the status of some of the switches
may not be consistent with the solution. For example, if a diode switch was assumed
to be off, and if the solution shows that
$V_p - V_n > V_{\mathrm{on}}$ for this diode, then the solution is not consistent.
The simulator would then try another switch on/off configuration, solve the new set
of equations, and check again for consistency.

We should point out that the example in Fig.~\ref{fig_sw_ex_1} is somewhat artificial~--
in real power electronic circuits, such a switch configuration would not be useful.
However, it serves as an effective vehicle to test the robustness of the solution
method, which is important for a general-purpose simulation package.

\subsection{Switch circuits: example 2}
\label{sec_sw_ex_2}
We will now consider the same circuit (Fig.~\ref{fig_sw_ex_1}) as in the previous
example but with a different switch on/off configuration, viz.,
$S_1$, $S_4$, $S_5$, $S_6$ on, and
$S_2$, $S_3$ off.
The ES and CTDC equations remain the same as before. In the following, we list only
the CTDV equations.

For the OFF branch (branch 3), we write equations relating the switch node voltages and 
$V_{sw}^{(k)}$, i.e.,
\begin{equation}
V_B - V_C - V_{sw}^{(2)} = 0,
\end{equation}
\begin{equation}
V_C - 0 - V_{sw}^{(3)} = 0.
\end{equation}

For branch 3, we equate the reverse bias for $S_2$ and $S_3$:
\begin{equation}
(V_{sw}^{(2)} - V_{\mathrm{on}}^{(2)}) -
(V_{sw}^{(3)} - V_{\mathrm{on}}^{(3)}) = 0.
\end{equation}

Next, for each ON branch with multiple switches, we equate their $i_{sw}$
variables. In our example, branch 4 is ON and has multiple switches, leading to
\begin{equation}
i_{sw}^{(5)} - i_{sw}^{(4)} = 0,
\end{equation}
\begin{equation}
i_{sw}^{(6)} - i_{sw}^{(4)} = 0.
\end{equation}

If there are ON branches which only contain switches and are in parallel,
we equate the branch currents. Branches 2 and 4 qualify, and we get
\begin{equation}
i_2 - i_4 = 0.
\end{equation}

Next, from the set of parallel ON branches, we take one branch (say, branch 2)
and write the KVL equations between the branch terminals. For branch 2, there
is only one KVL, viz.,
\begin{equation}
V_B - 0 - V_{sw}^{(1)} = 0.
\label{eq_sw_ex_2_1}
\end{equation}
For the remaining branches in the set~-- in this case, branch 4~-- we write KVL
for $N-1$ switches where $N$ is the total number of switches in branch 4.
For example, we may choose to write KVL for $S_5$ and $S_6$, and obtain
\begin{equation}
V_D - V_E - V_{sw}^{(5)} = 0,
\label{eq_sw_ex_2_2}
\end{equation}
\begin{equation}
V_E - 0 - V_{sw}^{(6)} = 0.
\label{eq_sw_ex_2_3}
\end{equation}
That completes the set of our CTDV equations, a total of 21 equations.

Note that, writing all $N$ KVL equations for branch 4 would make the equations
unsolvable. For branch 4, if we were to also write the KVL equation for $S_4$, i.e.,
\begin{equation}
V_B - V_D - V_{sw}^{(4)} = 0,
\label{eq_sw_ex_2_4}
\end{equation}
then we have, by adding
Eqs.~\ref{eq_sw_ex_2_2}-\ref{eq_sw_ex_2_4},
\begin{equation}
V_B = V_{sw}^{(4)} + V_{sw}^{(5)} + V_{sw}^{(6)},
\end{equation}
and that would clearly conflict with Eq.~\ref{eq_sw_ex_2_1}.

\section{Conclusions and future work}
\label{sec_conclusions}
To summarise, we have addressed singular matrix issues which arise in the
ELEX scheme presented in \cite{elex1} for using explicit methods to simulate
power electronic circuits. We have proposed topology-aware approaches to handle
circuits involving inductors and switches. For all examples discussed in this
paper, we have also verified that the proposed CTA approach gives the
expected results. For this purpose, individual programs were developed to
implement the ELEX-RKF scheme\,\cite{elex1} for each specific circuit.

The following future work is planned.
\begin{list}{\arabic{cntr1}.}{\usecounter{cntr1}}
 \item
  The ELEX scheme, along with the CTA approach described in this paper, will
  be implemented in the open-source package GSEIM\,\cite{gseimgithub}.
 \item
  Using GSEIM, comparison of simulation times will be carried out for several
  benchmark circuits, using implicit and explicit methods.
\end{list}
With both implicit and explicit options for electrical circuits as well as
ODE's, GSEIM is expected to become a useful open-source tool for a variety
of applications.

\bibliographystyle{IEEEtran}
\bibliography{ref4}

\end{document}